\documentclass[aps,prb,floatfix,twocolumn,showpacs,10pt]{revtex4-1}

\usepackage{epsfig,amsmath,amssymb,placeins,graphicx}

\begin{document}

\title{Spin Hot Spots in Single-Electron GaAs-based Quantum Dots}
\author{Martin Raith$^1$, Thomas Pangerl$^1$, Peter Stano$^{2,3}$, and Jaroslav Fabian$^1$}
\affiliation{$^1$Institute for Theoretical Physics, University of Regensburg, D-93040 Regensburg, Germany\\
$^2$RIKEN Center for Emergent Matter Science, 2-1 Hirosawa, Wako, Saitama, 351-0198 Japan\\
$^3$Institute of Physics, Slovak Academy of Sciences, 845 11 Bratislava, Slovakia}

\vskip1.5truecm

\begin{abstract}
Spin relaxation of a single electron in a weakly coupled double quantum dot is calculated numerically. The phonon assisted spin flip is allowed by the presence of the linear and cubic spin-orbit couplings and nuclear spins.
The rate is calculated as a function of the interdot coupling, the magnetic field strength and orientation, and the dot bias. 
In an in-plane magnetic field, the rate is strongly anisotropic with respect to the magnetic field orientation, due to the anisotropy of the spin-orbit interactions. The nuclear spin influence is negligible. In an out-of-plane field, the nuclear spins play a more important role due selection rules imposed on the spin-orbit couplings. Our theory shows a very good agreement with data measured in [Srinivasa, et al., PRL 110, 196803 (2013)], allowing us to extract information on the linear spin-orbit interactions strengths in that experiment. We estimate that they correspond to spin-orbit lengths of about 5-15 $\mu$m.
\end{abstract}


\maketitle

\section{Introduction}

Semiconductor heterostructure based quantum dots with confined electronic spins are among the most prominent platforms of spitronics\cite{zutic2004:RMP, fabian2007:APS} and quantum information related technology.\cite{loss1998:PRA, hanson2007:RMP, kloeffel2013:ARCMP, zwanenburg2013:RMP} The lifetime of information stored in a quantum dot spin qubit is limited by the spin relaxation\cite{elzerman2004:N,amasha2008:PRL} and decoherence.\cite{petta2005:S,koppens2008:PRL} Whereas the latter, mostly due to nuclear spins,\cite{khaetskii2002:PRL} can be suppressed by spin echo protocols,\cite{bluhm2010:N} the former is fundamentally limited by the relaxation through phonons.\cite{khaetskii2001:PRB, stavrou2006:PRB, stano2006:PRB, wu2010:PR}

Phonons do not couple to the electron spin directly. The spin relaxation is enabled by the spin-orbit interactions\cite{khaetskii2000:PRB, khaetskii2001:PRB, woods2002:PRB, alcalde2004:PE, alcalde2005:MJ, san-jose2006:PRL} or nuclear spins.\cite{erlingsson2001:PRB, khaetskii2002:PRL} Since these spin-dependent interactions are weak, compared to the confinement energy, the spin relaxation is very slow (may reach even seconds), which was one of the original motivations to consider spin qubits. The exception happens at points in the parameter space where levels (anti)cross. Here the spin relaxation rate is strongly enhanced, by orders of magnitude. Such points are called spin hot-spots.

The important influence that the spin hot-spots might imply on the spin relaxation was recognized in bulk metals\cite{fabian1998:PRL} and in quantum dots.\cite{golovach2004:PRL,stano2005:PRB} In the latter this influence is predicted to result in a very strong anisotropy in the spin lifetimes and the exchange interaction, which should be present generally, for various dot materials and charge occupations.\cite{stano2006:PRL, raith2011:PRB, raith2012:PRL, baruffa2010:PRL} However, it is only recently that spin hot spots were experimentally established in gated Si and GaAs quantum dots.\cite{yang2013:NC,srinivasa2013:PRL} 

Motivated by these recent experiments, here we investigate the spin relaxation in a single electron biased weakly coupled double dot in GaAs.\cite{cheng2004:PRB, sherman2005:PRB, florescu2006:PRB, romano2006:PRB, climente2006:PRB, semenov2007:PRB, jiang2008:PRB} This complements our studies of single electron unbiased double dots\cite{stano2005:PRB,stano2006:PRB,stano2006:PRL} and two electron biased double dots.\cite{raith2012:PRL} We investigate the relaxation rates anisotropy with respect to the in-plane magnetic field orientation, and compare the spin-orbit and nuclear fields effectiveness to induce the electron spin relaxation in in-plane and out-of-plane magnetic fields. We explain the observed relaxation rate intricate behavior by examining different channels that contribute to the total rate. Finally, we extract typical spin-orbit lengths in a GaAs quantum dot by fitting data from a recent experiment.\cite{srinivasa2013:PRL}

We organize the paper as follows. The model of the double dot, material parameters, and the numerical technique used for computation are outlined in Sec.~\ref{sec:model}. Sec.~\ref{sec:results} contains the numerical results for the relaxation rate for an in-plane magnetic field, and perpendicular field, comparison of the spin-orbit and nuclear effectiveness, and the fit of the experimental data from Ref.~\onlinecite{srinivasa2013:PRL}.

\section{Model\label{sec:model}}

\begin{figure}
\centerline{\psfig{file=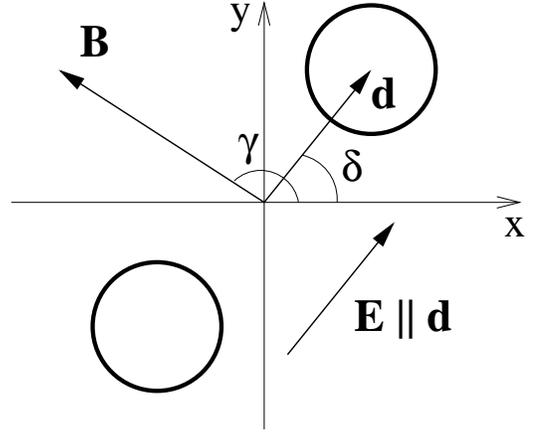,width=0.8\linewidth}}
\caption{The orientation of the potential dot minima (denoted as the two circles) with respect to the crystallographic axes ($x=[100]$ and $y=[010]$) is defined by the angle $\delta$. The magnetic field orientation is given by the angle $\gamma$. The electric field ${\bf E}$ is parallel to ${\bf d}$.}
\label{fig:definition}
\end{figure}

We consider a GaAs/AlGaAs heterojunction with growth direction $\mathbf{\hat{z}}=[001]$. The electrons at the interface are further confined by the electrostatic field of top gates. Using the envelope function approximation, the two-dimensional Hamiltonian of a single electron in a biased double dot reads as
\begin{equation}
 H=T+V+H_{\mathrm{Z}}+H_{\mathrm{so}}+H_{\mathrm{nuc}}.
\label{hamiltonian}
\end{equation}
Here $T=\mathbf{P}^2/2m$ is the kinetic energy with the electron effective mass $m$, and the kinematic momentum $\mathbf{P}=-i \hbar \boldsymbol{\nabla}+e\mathbf{A}$, where $e$ is the proton charge.
The two dimensional vector potential reads $\mathbf{A}=-\left(yB_z/2\right)\hat{\mathbf{x}}+\left(xB_z/2\right)\hat{\mathbf{y}}$, where $\hat{\mathbf{x}}=[100]$ and $\hat{\mathbf{y}}=[010]$. The magnetic field is
$\mathbf{B}=\left(B_{\parallel}\cos\gamma,B_{\parallel}\sin\gamma,B_z\right)$, where $\gamma$ is the angle between the in-plane component of the magnetic field and the $[100]$-direction. The orbital effects of the in-plane magnetic field are neglected.\cite{stano2006:PRB} The in-plane position vector is $\mathbf{r}=\left(x,y\right)$. The double dot is defined by the bi-quadratic confinement potential,\cite{helle2005:PRB,pedersen2007:PRB,li2010:PRB}
\begin{equation}
 V=\frac{\hbar^2}{2ml_0^4} \mathrm{min} \left\{\left(\mathbf{r}-\mathbf{d}\right)^2,\left(\mathbf{r}+\mathbf{d}\right)^2 \right\}+e\mathbf{E}\cdot\mathbf{r}.
\label{confining_potential}
\end{equation}
For zero electric field ${\bf E}$ the potential minima are located at $\pm\mathbf{d}$, and we call $2d/l_0$ the (dimensionless) interdot distance. The angle between ${\bf d}$ and [100] is denoted as $\delta$. The potential strength is characterized by the confinement energy $E_0=\hbar^2/ ml_0^2$, with the confinement length $l_0$. The electric field $\mathbf{E}$ applied along $\mathbf{d}$ leads to an energy offset between the potential minima, $\epsilon=2eEd$, which we call bias in further. The geometry is summarized in Fig.~\ref{fig:definition}.

The Zeeman term reads
\begin{equation}
 H_{\mathrm{Z}}=\frac{g}{2} \mu_B \mathbf{B} \cdot \boldsymbol{\sigma},
\label{zeeman_hamiltonian}
\end{equation}
where $g$ is the effective conduction band $g$ factor, $\mu_B$ is the Bohr magneton, and $\boldsymbol{\sigma}$ is the vector of the Pauli matrices. 

The spin-orbit coupling, $H_{\mathrm{so}}=H_{\mathrm{br}}+H_{\mathrm{d}}+H_{\mathrm{d3}}$, consists of three terms, the Bychkov-Rashba, the linear, and the cubic Dresselhaus spin-orbit coupling.\cite{zutic2004:RMP,fabian2007:APS}  The Bychkov-Rashba Hamiltonian, arising from the heterostructure asymmetry, reads as\cite{bychkov1984:JPC}
\begin{equation}
 H_{\mathrm{br}}=\frac{\hbar}{2ml_{\mathrm{br}}} \left(\sigma_x P_y-\sigma_y P_x\right),
\label{bychkov-rashba}
\end{equation}                                                                                                                                                                                                            
where the strength is parameterized by the spin-orbit length $l_{\mathrm{br}}$. The bulk inversion asymmetry of the zinc-blende structure enables the Dresselhaus interaction.\cite{dresselhaus1955:PR} It consists of two terms: linear, and cubic (referring to the power of the momentum operator),
\begin{align}
H_{\mathrm{d}}&=\frac{\hbar}{2ml_{\mathrm{d}}}\left(-\sigma_x P_x+\sigma_y P_y\right), 
\label{dresselhaus}\\ 
H_{\mathrm{d3}}&=\frac{\gamma_c}{2\hbar^3}\left(\sigma_x P_x P_y^2-\sigma_y P_y P_x^2\right)+\mathrm{H.c.},
\label{dresselhaus3}
\end{align}
respectively. The linear term is parameterized by the spin-orbit length $l_{\mathrm{d}}$, and $\gamma_{\mathrm{c}}$ is a material parameter. 

The last term in Eq.~(\ref{hamiltonian}) describes the hyperfine interaction of the confined electron with the lattice's nuclei,\cite{schliemann2003:JPCM,stano2013:PRB}
\begin{equation}
H_{\mathrm{nuc}}=\beta\sum\limits_{n}\mathbf{I}_n\cdot\boldsymbol{\sigma}\delta\left(\mathbf{R}-\mathbf{R}_n\right),
\label{hyperfine}
\end{equation}
where $\beta$ is a constant, and $\mathbf{I}_n$ and $\mathbf{R}_n$ are the spin and the position of the n-th nucleus. Here the vectors of position are three-dimensional, $\mathbf{R}=(\mathbf{r},z)$. The electron wavefunction along the growth direction, $\Psi(z)$, defines an effective width $h_z=\left(\int\mathrm{d}z\lvert\Psi(z)\rvert^4\right)^{-1}$.\cite{erlingsson2002:PRB} We assume $\Psi(z)$ to be the ground state of a hard-wall confinement of width $w$, and get $h_z=2w/3$. 

The relaxation is enabled by acoustic phonons. The electron-phonon interaction Hamiltonian reads as
\begin{equation}\label{phonon_potentials}
 H_{\mathrm{ph}}=i\sum\limits_{\mathbf{Q},\lambda}\sqrt{\frac{\hbar Q}{2\rho V c_\lambda}}V_{\mathbf{Q},\lambda}\left(b_{\mathbf{Q},\lambda}^{\dagger}e^{i\mathbf{Q}\cdot\mathbf{R}}-
b_{\mathbf{Q},\lambda}e^{-i\mathbf{Q}\cdot\mathbf{R}}\right) ,
\end{equation}
with $\lambda=l,t1,t2$ denoting the polarization of the phonons (one longitudinal and two transverse). The three-dimensional phonon wave vector is $\mathbf{Q}$. The phonon creation and annihilation operator is given by $b$ and $b^{\dagger}$, respectively. The mass density of the crystal is $\rho$, its volume is $V$, and the sound velocities are $c_{\lambda}$. The deformation potential is $V_{\mathbf{Q},\lambda}=\sigma_e \delta_{\lambda,l}$ and the piezoelectric potential is $V_{\mathbf{Q},\lambda}=-ieh_{14}N_{\lambda}/Q^3$ with $N_{\lambda}=2\left(q_xq_y\hat{e}^{\lambda}_z+q_zq_x\hat{e}^{\lambda}_y+q_yq_z\hat{e}^{\lambda}_x\right)$. The unit polarization vector is $\mathbf{\hat{e}^{\lambda}}$.

The relaxation rate for the transition from state $\left|i\right\rangle$ to $\left|f\right\rangle$ is calculated using the Fermi's Golden Rule in the zero temperature limit,
\begin{equation}
 \Gamma_{if}=\frac{\pi}{\rho V}\sum\limits_{\mathbf{Q},\lambda}\frac{Q}{c_{\lambda}}|V_{\mathbf{Q},\lambda}|^2|M_{if}|^2\delta\left(E_{if}-E_{\mathbf{Q}}\right),
 \label{eq:relax_rate}
\end{equation}
where $M_{if}=\left\langle i\left|e^{i\mathbf{Q}\cdot\mathbf{R}}\right| f\right\rangle$ is the transition matrix element, and $E_{if}$ is the energy difference between $\left|i\right\rangle$ and $\left|f\right\rangle$. To incorporate nuclei, we average the relaxation rate in Eq.~\eqref{eq:relax_rate} over several (typically 50) random configurations of an unpolarized nuclear bath---see Ref.~\onlinecite{raith2012:PRL} for details.

In numerics we use the material parameters of bulk GaAs: $m=0.067m_e$, where $m_e$ is the free electron mass, $g=-0.44$, $\rho=5300$ kg/$\mathrm{m}^3$, $c_l=5290$ m/s, $c_t=2480$ m/s, $\gamma_c=27.5$ eV$\mathring{\mathrm{A}}^3$, $\sigma_e=7$ eV, $eh_{14}=1.4\times10^9$ eV/m, $\beta=1\mathrm{\mu}\mathrm{eVnm}^3$, and $I=3/2$.
The quantum dot parameters are $l_0=34$\,nm ($E_0=1$\,meV), $l_{\mathrm{br}}=2.42$ $\mathrm{\mu}$m and $l_{\mathrm{d}}=0.63$ $\mu$m.\cite{footnote8} We use the coupling strength of $2d/l_0=4.35$, corresponding to a tunneling energy of $t=0.01$\,meV. The orientation of the dots is along [110], i.e. $\delta=45^\circ$, unless stated otherwise. 

Since the energy spectrum of the Hamiltonian in Eq.~\eqref{hamiltonian} cannot be solved for analytically, we treat it numerically using the finite differences method with Dirichlet boundary conditions\cite{Cheng2005:EABE} including the magnetic field via the Peierl's phase.\cite{peierls1933:ZfP} The resulting eigenvalue problem is then solved using the Lanczos algorithm.\cite{Lanczos1950:JRNBS} In the numerics we use grid dimensions of typically around $200\times200$ grid points. The relative error is below $10^{-5}$.

\begin{figure}
 \centering
 \includegraphics[width=\linewidth]{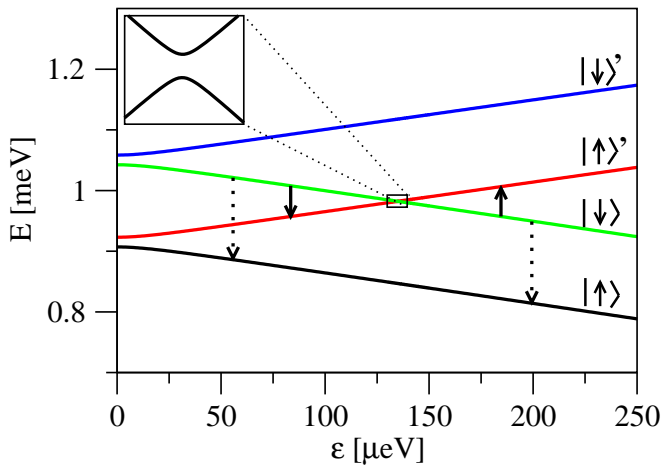}
\caption{\label{fig:ani_spectrum} (Color online) Calculated energy spectrum of a GaAs double dot as a function of the bias $\epsilon$ in an in-plane magnetic field $B$ of 7\,T with $\delta=\gamma=45^{\circ}$ and $t=0.01$ meV. The states are labeled according to their spin orientation and parity at $\epsilon=0$ (unprimed for even, primed for odd). The inset magnifies the anticrossing. The thin arrows denote transitions that contribute to the measured relaxation rate.}
\end{figure}

\section{Results}

\label{sec:results}

\subsection{In-plane magnetic field anisotropy\label{sec:results:in-plane}}

Let us first look at the dot energy spectrum. Figure~\ref{fig:ani_spectrum} shows the lowest four levels as a function of the bias for a weakly coupled double dot. States are denoted according to their spatial inversion parity at zero bias: states with a prime have odd, and states without a prime have even parity. At a detuning energy of about the Zeeman energy, $\epsilon=0.178$\,meV, the states $\left|\uparrow\right\rangle'$ and $\left|\downarrow\right\rangle$ form an anisotropic anticrossing due to spin-orbit coupling. The anticrossing energy is maximal for an orientation of $\gamma=45^{\circ}$, and absent if $\gamma=135^{\circ}$. This special point in the spectrum is the spin hot spot.\cite{fabian1998:PRL,stano2005:PRB} Here the spin orientation smoothly changes from an up to a down state and vice versa. 

\begin{figure}
 \centering
 \includegraphics[width=\linewidth]{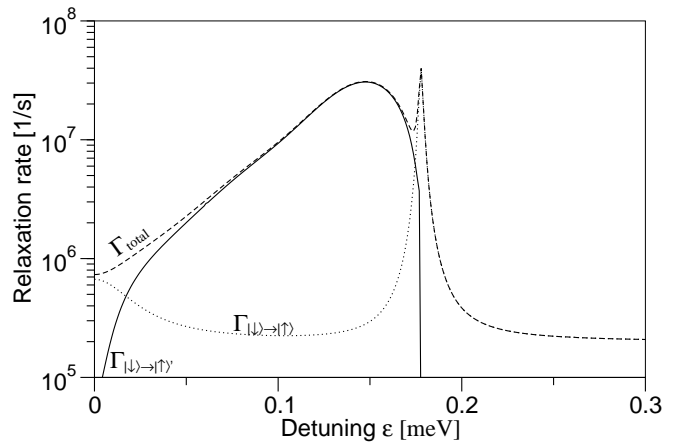}
\caption{\label{fig:channels}Calculated spin relaxation rate, resolved into channels, of a double dot as a function of detuning for $\delta=\gamma=45^{\circ}$ with $B=7$\,T, $t=0.01$ meV and $T=0$ K. The hyperfine coupling is neglected. The dotted, solid, and dashed line gives $\Gamma_{|\downarrow\rangle\to | \uparrow\rangle}$, $\Gamma_{|\downarrow\rangle\to | \uparrow^\prime\rangle}$, and $\Gamma$, respectively.}
\end{figure}

We define the "spin relaxation rate" $\Gamma$ according to what is measured in corresponding experiments.\cite{srinivasa2013:PRL} The initial state for the transition is the lowest spin down state, $\left|\downarrow\right\rangle$, while the transition is considered as completed if the lowest state, $\left|\uparrow\right\rangle$, is detected. Since the transitions between spin alike states are much faster than a duration of the measurement cycle, they can be considered instantaneous and we have
\begin{equation}
\Gamma \approx \Gamma_{|\downarrow\rangle\to | \uparrow\rangle} + \Gamma_{|\downarrow\rangle\to | \uparrow^\prime\rangle}.
\label{eq:gamma definition}
\end{equation}
The individual transition rates for a weakly coupled double dot are plotted as a function of the bias in Fig.~\ref{fig:channels}. The relaxation rate between the two lowest Zeeman split states $\Gamma_{\left|\downarrow\right\rangle\rightarrow\left|\uparrow\right\rangle}$ is, apart from the anticrossing, not varying much, due to the energy difference being constant. On the other hand, the transition into the first excited state $\left|\uparrow\right\rangle'$ is highly non-monotonic. Initially it grows, since, as the two states become closer in energy, it is easier to admix the spin-opposite component into the states. If the energy difference becomes too small, the rate drops, as now the diminishing density of states of phonons takes over the trend. For detunings beyond the anticrossing, the second term of the right hand side of Eq.~\eqref{eq:gamma definition} will be suppressed at low temperatures, which contributes to the strong asymmetry of the relaxation rates as a function of the bias with respect to the position of the anticrossing.

\begin{figure}
 \centering
 \includegraphics[width=\linewidth]{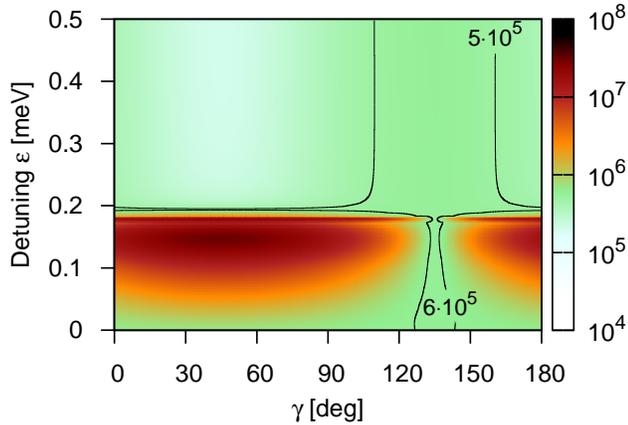}
\caption{\label{fig:anisotropy_detuning}(Color online) Calculated spin relaxation of a double dot as a function of detuning and orientation of the in-plane magnetic field $B=7$\,T, with $\delta=45^\circ$, $t=0.01$ meV and $T=0$ K. The hyperfine coupling is neglected. The corresponding energy spectrum (at $\gamma=45^{\circ}$) is shown in Fig.~\ref{fig:ani_spectrum}. The rate is plotted according to the color scale on the right in inverse seconds. The labeled contours represent equirelaxation lines.}
\end{figure}

The spin-orbit enabled relaxation rate as a function of detuning and orientation of an in-plane magnetic field is plotted in Fig.~\ref{fig:anisotropy_detuning}. It shows the anisotropic relaxation landscape and the existence of two principal axes for the in-plane magnetic field orientation: parallel ($\gamma=45^{\circ}$) and perpendicular ($\gamma=135^{\circ}$) to the dot main axis $\mathbf{d}$. For small detunings, and in the vicinity of the spin hot spot at $\epsilon=0.178$\,meV, the relaxation rate is strongly suppressed if $\gamma=135^{\circ}$. On the other hand, the relaxation rate for large detunings is minimal if $\gamma=45^{\circ}$, as here the system has single dot character. This directional switch of the axis of minimal relaxation has previously been found in two-electron double dots\cite{raith2012:PRL,raith2012:PRB} and can be understood from the effective, spin-orbit induced, magnetic field.\cite{stano2006:PRL} 
It is only for $\gamma=135^{\circ}$ that changing between unbiased and highly biased configurations can be achieved without passing through a regime of strongly enhanced spin relaxation. This feature is known as an easy passage.\cite{stano2006:PRL}

\subsection{Spin-orbit vs nuclear fields}

Let us now comment on the role of nuclei and how the presented results are altered in the presence of hyperfine-induced spin relaxation. For the double quantum dot considered above, we find that the relaxation rates due to the nuclei are typically 2 orders of magnitude smaller than the rates given in Fig.~\ref{fig:anisotropy_detuning}. The exception occurs at the spectral anticrossing ($\epsilon\approx0.178$ meV) because the states $\left|\downarrow\right\rangle$ and $\left|\uparrow\right\rangle'$ are always coupled by the nuclear spins irrespective of the orientation of the in-plane magnetic field. Thus, the hyperfine-induced spin relaxation becomes dominant at the anticrossing along the easy passage, where the spin-orbit contribution to the relaxation is of the order of $10^5$\,s$^{-1}$. However, the impact here is rather small, as we show now.

\begin{figure}
 \centering
 \includegraphics[width=\linewidth]{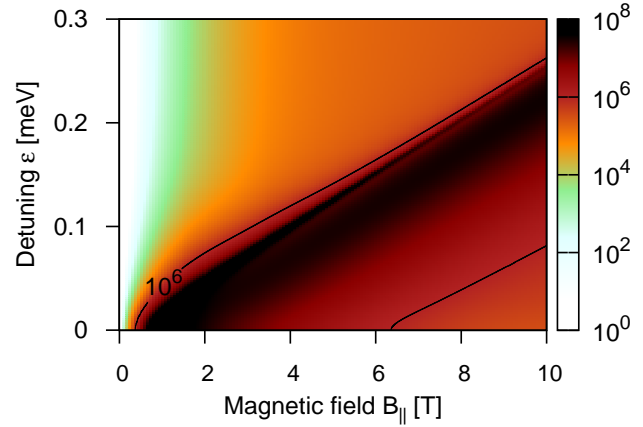}
\caption{\label{fig:detuning_soc_45}(Color online) Calculated spin relaxation of a double dot as a function of detuning and magnitude of the in-plane magnetic field with $t=0.01$\,meV, $B=7$\,T, $\delta=\gamma=45^{\circ}$, and $T=0$ K. The hyperfine coupling is neglected. The rate is plotted according to the color scale on the right in inverse seconds. The labeled contours represent equirelaxation lines.}
\end{figure}
\begin{figure}
 \centering
 \includegraphics[width=\linewidth]{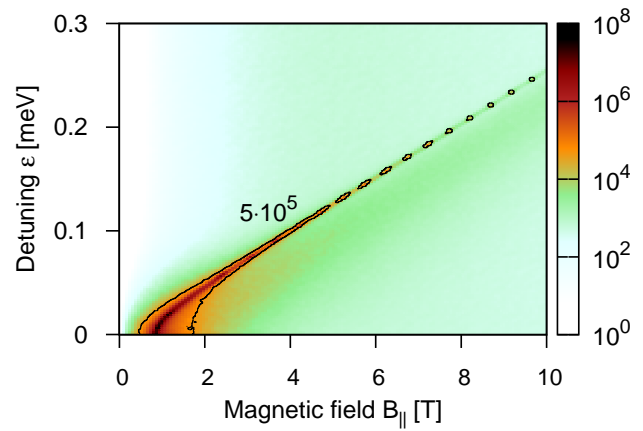}
\caption{\label{fig:detuning_hypf_45}(Color online) Same as Fig.~\ref{fig:detuning_soc_45}, but now only hyperfine coupling is considered (no spin-orbit coupling). The cellular structure of the plot around the anticrossing is a finite resolution artifact, not a physical effect.}
\end{figure}

We present the spin relaxation rates enabled by either spin-orbit coupling or hyperfine coupling for the double dot with parameters given above in Fig.~\ref{fig:detuning_soc_45} and Fig.~\ref{fig:detuning_hypf_45}, respectively. The in-plane magnetic field orientation for this comparison is chosen to be $\gamma=45^{\circ}$, i.e.~away from the easy passage. We see that the spin-orbit contribution is dominant over the whole parameter range. The spike in the relaxation rate map of Fig.~\ref{fig:detuning_hypf_45} becomes relevant only in the easy passage configuration (or for magnetic fields below 2\,T). However, we find that the impact on the total relaxation rate is rather weak because of the small width of the spike. Particularly, the spike is hardly visible for magnetic fields of 6\,T or more.


\subsection{Perpendicular magnetic field\label{sec:results:perpendicular}}
For completeness, we now consider a symmetric double quantum dot in an external magnetic field perpendicular to the dot plane. We focus on the dependence of the spin relaxation on the magnetic field magnitude and the interdot distance, and compare the impact of spin-orbit and hyperfine coupling on the relaxation rates. For more on biased dots in perpendicular magnetic fields, see Ref.~\onlinecite{Pangerl2012}.

Perpendicular magnetic field has also orbital effects, resulting in an effective confinement length\cite{stano2005:PRB} $l_{\mathrm{B}}=\left(l_0^{-4}+B^2e^2/4\hbar^2\right)^{-1/4}$. This effectively changes the interdot coupling, and the tunneling energy decreases. The positions of the level crossings in the energy spectrum are therefore strongly dependent on both the interdot distance and the magnetic field strength.

\begin{figure}
\flushleft
 \includegraphics[trim=1cm 0cm 0cm 0cm, clip=true, width=\linewidth]{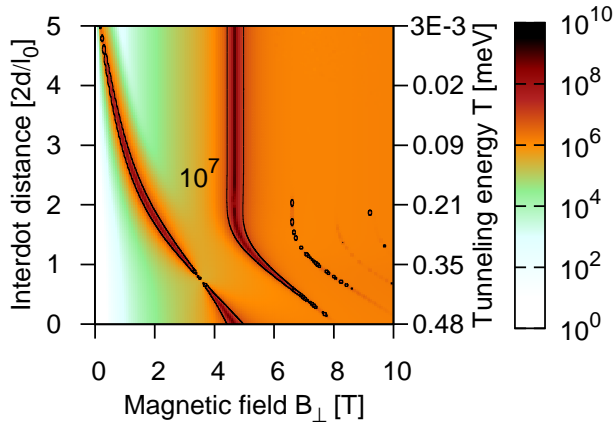}
\caption{\label{fig:interdot:soc}(Color online) Spin relaxation of a double dot as a function of perpendicular magnetic field and interdot coupling. The rate is given in $\mathrm{s}^{-1}$. The solid lines represent equirelaxation lines. The relaxation rate was calculated considering only the spin-orbit coupling.}
\end{figure}

\begin{figure}
\flushleft
 \includegraphics[trim=1cm 0cm 0cm 0cm, clip=true, width=\linewidth]{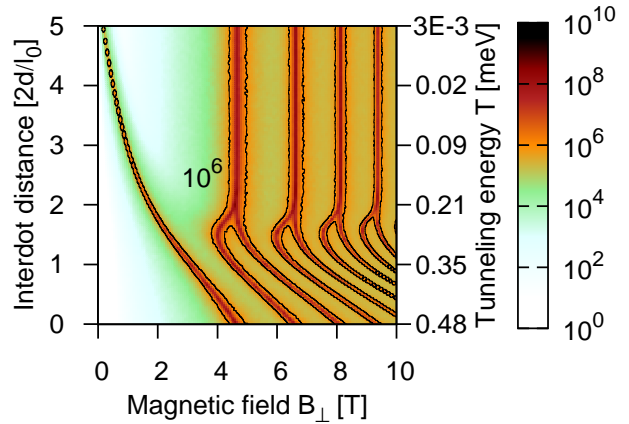}
\caption{\label{fig:interdot:hypf}(Color online) Spin relaxation of a double dot as a function of perpendicular magnetic field and interdot coupling. The rate is given in $\mathrm{s}^{-1}$. The solid lines represent equirelaxation lines. The relaxation rate was calculated considering only the hyperfine coupling.}
\end{figure}

We plot the spin relaxation rates of a double dot in a perpendicular magnetic field for the cases of either spin-orbit coupling or hyperfine coupling in Figs.~\ref{fig:interdot:soc} and \ref{fig:interdot:hypf}, respectively. Without the nuclear field (Fig.~\ref{fig:interdot:soc}), the first hot spots of the single dot ($d=0$) are at $B\approx4.5$\,T, $B\approx7.9$\,T, etc. We find that the spikes in the relaxation rate map in Fig.~\ref{fig:interdot:soc} generally become less pronounced for stronger magnetic fields. Except for the first anticrossing at $B\approx4.5$\,T, the level crossings for interdot distances $2d/l_0\gtrsim2$ ($T\lesssim0.2$\,meV) are found at a constant magnetic field.

Switching off the spin-orbit coupling and considering only the coupling to the nuclei (Fig.~\ref{fig:interdot:hypf}), we find more spikes in the relaxation rate map as compared to Fig.~\ref{fig:interdot:soc}. The difference between the influence of nuclei and spin-orbit coupling is due to the chosen confinement profile. Namely, in a parabolic well, the linear-in-$p$ spin-orbit interactions couple only Fock-Darwin states with orbital momentum differing by 1.\cite{destefani2004:PRB} This leads to a very strong suppression of the width of higher anticrossings induced by the spin-orbit interaction. On the other hand, there is no such selection rule for unpolarized nuclei and in this case the widths of consecutive anticrossings decay much slower. Since the parabolic confinement is believed to be a good description of the low lying part of the spectrum, we conclude that the hyperfine-induced spin relaxation plays a more important role if the external magnetic field is perpendicular.

\subsection{Extracting spin-orbit lengths from Ref.~\onlinecite{srinivasa2013:PRL}}

Having available a quantitatively faithful theory, we fit the data measured in Ref.~\onlinecite{srinivasa2013:PRL}, an experiment on the spin relaxation in a single electron weakly coupled GaAs double dot. We aim at extraction of the spin-orbit lengths. Despite their crucial importance for spintronics applications and theory, their values are not reliably established in small (occupied by few electrons) quantum dots, where the strong confinement may renormalize the values extrapolated from measurements in quantum wells or bulk.

Ref.~\onlinecite{srinivasa2013:PRL} gives the following experimentally accessible parameters: the confinement energy of 1 meV, the tunneling energy of 8 $\mu$eV, the magnetic field of 6.5 T, applied along the dot main axis, $\gamma=\delta$, and the anticrossing occurring at the detuning of 0.136 meV. They translate into the following parameters of our model: $l_0=34$ nm, $g=-0.364$, and $d=76.5$ nm. For $m$, $\gamma_c$ and phonon characteristics we use the bulk values, as given below Eq.\eqref{eq:relax_rate}. Finally, we use a finite temperature of 0.25 K, and neglect nuclear spins. 

\begin{figure}
\flushleft
 \includegraphics[width=\linewidth]{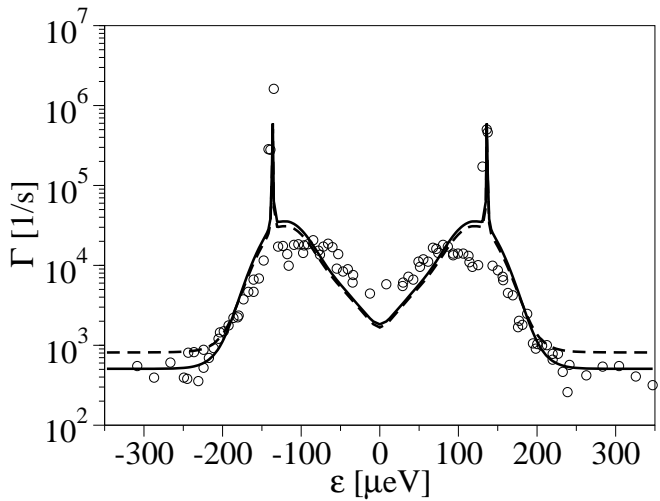}
\caption{\label{fig:fitting}Comparison of calculated (solid line; parameters from the first line of Tab.~\ref{tab:fits}) spin relaxation rates to the ones measured in Ref.~\onlinecite{srinivasa2013:PRL} (symbols). The dashed line is the result of the minimization with a fixed dot orientation, $\delta=45^\circ=\gamma$, which gave $\alpha=0.84$ meV\AA, and $\beta=0.47$ meV\AA.}
\end{figure}

We keep the spin-orbit lengths $l_{br}$, $l_d$, and the dot orientation $\delta$ (the angle between the dot main axis and the crystallographic axis [100]) as fitting parameters. We adopt a standard procedure\cite{numrec} and fit by minimizing the $\chi^2$ measure
\begin{equation}
\chi^2 = \sum_{\epsilon_i} \left(\log[\Gamma^\star(\epsilon_i)] - \log[\Gamma(\epsilon_i,l_{br},l_d,\gamma)] \right)^2.
\label{eq:chi}
\end{equation}
Here $\epsilon_i$ labels different measurements, and $\Gamma^\star$ are measured values (100 data points). Since the rates vary over orders of magnitude, we use a logarithmic scale.

Because of a highly non-linear shape of the relaxation rate curves, the figure of merit of the fit, the function $\chi^2$, has many local minima in the fitting parameters space. We give several examples in Tab.~\ref{tab:fits}. The rather small differences in values of $\chi^2$ in these minima mean that all these parameter sets fit the data almost equally well. This also gives a very crude estimate on the reliability of the extracted values of the spin-orbit strengths---their relative sign remains unknown and their magnitudes can not be established better than within a factor of 3. The fit corresponding to the parameters in the first line of the table is plotted together with the measured data in Fig.~\ref{fig:fitting}. We also plot a result of minimization with a fixed dot orientation, $\delta = 45^\circ=\gamma$, which might have been the case in the experiment.\footnote{V.~Srinivasa (private communication).} This would also make the values in lines  2 and 3 of the Tab.~I more probable than others. 

\begin{table}
\begin{tabular}{ccccccc}
\hline\hline
\,\,set \,\,& \,$\delta$\, & \,\,$\alpha$[meVA]\,\, & \,\,$\beta$[meV\AA]\,\, & \,\,$l_{br}$[$\mu$m]\,\, & \,\,$l_d$[$\mu$m]\,\, & \,$\chi^2_{\rm min}$\,\\
\hline
1 & 307$^\circ$ & -1.34 & 1.51 & -4.2  & 3.8  & 9.66\\
2 & 60$^\circ$ & 0.65 & 0.89 & 8.8  & 6.4  & 9.69\\
3 & 62$^\circ$ & 0.55 & 0.82 & 10.3  & 7.0  & 9.71\\
4 & 203$^\circ$ & 0.34 & 0.67 & 16.7  & 8.4  & 9.76\\
5 & 294$^\circ$ & 0.48 & -0.45 & 11.8  & -12.6  & 9.87\\
\hline\hline
\end{tabular}
\caption{Fitted spin-orbit lengths $l_{br}$ and $l_d$ and the dot orientation $\delta$ (angle between the main dot axis and [100]). Each set corresponds to a local minimum of $\chi^2$, Eq.~\eqref{eq:chi}, in the parameter space. The spin-orbit lengths are also given in alternative units through $\alpha=\hbar^2/2 m l_{br}$ and $\beta=\hbar^2/2 m l_d$. Our definition of the spin-orbit lengths $l_{so}$ in Eqs.~ \eqref{bychkov-rashba} and \eqref{dresselhaus} is such that in a one dimensional model these Hamiltonians induce a rotation of the spin by an angle $2\pi r /l_{so}$ upon a spatial displacement of the electron by a distance $r$.}
\label{tab:fits}
\end{table}

\section{Summary\label{sec:conclusion}}
We have calculated phonon induced spin relaxation rates enabled via spin-orbit coupling and hyperfine coupling of single electron states in biased double quantum dots. 
We find strong anisotropies in the relaxation rate, due to anisotropy of the underlying spin-orbit interactions, and the related spin hots and easy passages, known from works on unbiased dots. For the spin-orbit strengths of the order of 1 $\mu$m, chosen by fitting  data measured in Ref.~\onlinecite{elzerman2004:N}, we find that the contribution of nuclear spins is negligible. Fitting data from a different experiment of Ref.~\onlinecite{srinivasa2013:PRL}, we extract the spin-orbit lengths of the order of 10 $\mu$m, for which nuclear spin contribution is roughly comparable to that of the spin-orbit interactions. To nail down the spin-orbit interactions strengths with better confidence, the measurement of the rate as a function of the magnetic field orientation is called for. In addition, such a measurement is ideal for separate identification of the two linear spin-orbit strengths.



\acknowledgments
This work was supported by DFG under grant SPP 1285 and SFB 689. P.S. acknowledges support from meta-QUTE ITMS NFP 26240120022, CE SAS QUTE, and COQI APVV-0646-10.

\bibliography{../../references/quantum_dot}

\end{document}